\PassOptionsToPackage{english}{babel}
\documentclass[%
 reprint,
superscriptaddress,
 amsmath,amssymb,
 aps,abbrv,
prb,
floatfix,
]{revtex4-1}

\usepackage[english]{babel}

\usepackage{graphicx}
\usepackage{dcolumn}
\usepackage{bm}
\usepackage{hyperref}
\usepackage[mathlines]{lineno}

\begin{document}

\preprint{APS/123-QED}

\title{Graded index lenses for spin wave steering}

\author{N. J. Whitehead}
\email{nw276@exeter.ac.uk}
\affiliation{Department of Physics \& Astronomy, University of Exeter, Stocker Road, Exeter, EX4 4QL, UK.}
\author{S. A. R. Horsley}
\affiliation{Department of Physics \& Astronomy, University of Exeter, Stocker Road, Exeter, EX4 4QL, UK.}
\author{T. G. Philbin}
\affiliation{Department of Physics \& Astronomy, University of Exeter, Stocker Road, Exeter, EX4 4QL, UK.}
\author{V. V. Kruglyak}
\email{V.V.Kruglyak@exeter.ac.uk}
\affiliation{Department of Physics \& Astronomy, University of Exeter, Stocker Road, Exeter, EX4 4QL, UK.}

\date{\today}

\begin{abstract}
We use micromagnetic modelling to demonstrate the operation of graded index lenses designed to steer forward-volume magnetostatic spin waves by 90 and 180 degrees. The graded index profiles require the refractive index to diverge in the lens center, which, for spin waves, can be achieved by modulating the saturation magnetization or external magnetic field in a ferromagnetic film by a small amount. We also show how the 90$^\circ$ lens may be used as a beam divider. Finally, we analyse the robustness of the lenses to deviations from their ideal profiles.

\end{abstract}

\maketitle


\section{Introduction}

Future wave-based computers will need to carry out certain functions to control propagating waves. One important function is to steer a wave beam or a wave packet in a controlled manner. For spin waves in ferromagnets \cite{stancil_spin_2009-1, kruglyak_magnonics_2010, lenk_building_2011, chumak_magnon_2015}, a possible contender for wave-based computing, this has mostly been investigated in terms of confining waves along curved waveguides \cite{vogt_spin_2012, xing_how_2013, xing_excitation_2015, garcia-sanchez_narrow_2015, lan_spin-wave_2015, sadovnikov_spin_2017}. However, these waveguides may suffer from losses/scattering in bends, and usually have a large spatial footprint. An alternative solution is to steer spin waves via a graded refractive index \cite{davies_graded-index_2015, dzyapko_reconfigurable_2016, gruszecki_spin-wave_2018, vogel_control_2018}, which smoothly alters the wave trajectory with minimal reflections \cite{whitehead_luneburg_2018}. To achieve a graded index for spin waves, one must gradually change a magnonic parameter on a length scale much smaller than the wavelength.

In optics, wave steering via a graded index is a well-established technique. One spatially-efficient method of steering is via rotationally-symmetric profiles (lenses), which are specifically designed to steer light by a certain angle between 0$^\circ$ and 360$^\circ$, and can do so from any direction of incidence \cite{eaton_spherically_1952, cornbleet_generalised_1981, minano_perfect_2006, schmiele_designing_2010, chang_enhanced_2012, sarbort_spherical_2012}. Although these lenses are designed to work with light, the same analysis applies to any other wave, supposing that the dispersion relation is known.

A practical problem with these lenses is that they require a singular refractive index in the center, while even a moderately large refractive index is difficult to achieve in most areas of wave physics, One technique to avoid this problem is via transformation optics \cite{tyc_transmutation_2008, hooper_transmutation_2013,horsley_removing_2014-1}. The profile can also be truncated, but this often results in an incorrect trajectory \cite{zentgraf_plasmonic_2011}. 
\begin{figure}
\centering
\includegraphics[width=\linewidth]{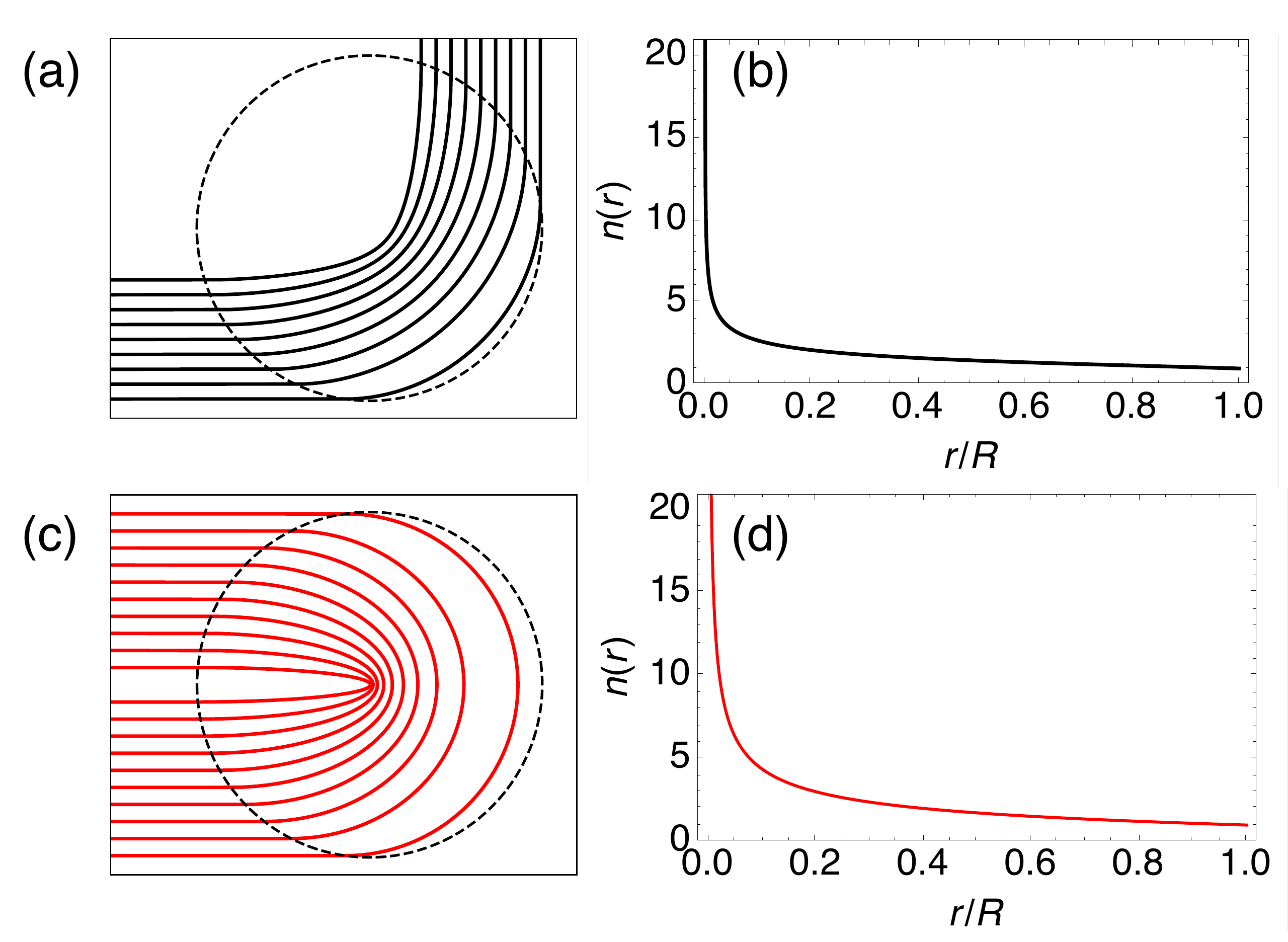}
\caption{\label{fig:index} Images (a) and (c) show the ray paths, and images (b) and (d) give the refractive index profile for the 90$^\circ$ and Eaton lenses (lens radius $R$), respectively.}
\end{figure}

Here, we show that an extremely high refractive index can quite easily be achieved for magnetostatic (dipolar) spin waves in the forward-volume geometry. Although a singular index is obviously still impossible, the refractive index can become high enough to closely match the required refractive index profile of these steering lenses. We use micromagnetic modelling to demonstrate how two of these lenses can be realized for spin waves in the dipolar regime, and analyse the lenses' robustness to profile deviations. 

\section{Theory of Spin Wave Steering Lenses}

We first explain the properties of the steering lenses, and then show how they may be implemented for spin waves. We will be using the 90$^\circ$ lens \cite{cornbleet_generalised_1981, schmiele_designing_2010} and Eaton (180$^\circ$) lens \cite{eaton_spherically_1952}. Fig. \ref{fig:index} compares their respective refractive index profiles, defined as \cite{sarbort_spherical_2012}
\begin{equation}
\label{eq:90}
\text{90$^\circ$ lens:} \quad (r/R) n^4 - 2n + (r/R) = 0,
\end{equation}
\begin{equation}
\label{eq:Eaton}
\text{180$^\circ$ (Eaton) lens:} \quad n(r) = \sqrt{\frac{2}{(r/R)} - 1},
\end{equation}
where $r$ is the radial coordinate and $R$ is the radius of the lens in each case. Note that the profile for the 90$^\circ$ lens \eqref{eq:90} is defined implicitly here. 
 
Defining a refractive index profile for spin waves is non-trivial, since the dispersion relation is strongly dependent on the geometry, and is always nonlinear. In some geometries it is also highly anisotropic. The simplest way to implement the rotationally-symmetric profiles is via a geometry with an isotropic dispersion relation, and design each lens for a fixed incident wave frequency, although it should work also for a wave packet with a small frequency spread. The refractive index is defined as the ratio of the wave number inside the lens, $k(r)$, to that outside the lens, $k_\text{ref}$, 
\begin{equation}
\label{eq:luneburg_sw}
n(r) = \frac{k(r)}{k_\text{ref}}.
\end{equation}
To change the wave number and thus the index for the given wave frequency, we need to change the dispersion relation by varying one of the bulk material parameters, or film thickness \cite{davies_graded-index_2015, dzyapko_reconfigurable_2016, gruszecki_spin-wave_2018, vogel_control_2018, whitehead_luneburg_2018}. 

We then need to choose an isotropic dispersion relation that enables a large change in $k$, and thus $n$. This requirement is satisfied in the dipolar-dominated regime, in the forward-volume geometry, where the magnetization is directed normal to the film plane. The dipole-dipole interaction dominates the dispersion for spin wave wavelengths $\lambda$ of millimeters to micrometers. At the other end of the spectrum, the short-range exchange interaction dominates, for wavelengths from tens to hundreds of nanometers. In the crossover regime, the dispersion curve flattens out before the exchange interaction begins to have a stronger influence. It is this shallow gradient in the crossover regime that enables a large index to be obtained. The forward-volume dipole-exchange dispersion relation can be written for the angular frequency $\omega(k)$ as \cite{kalinikos_theory_1986}
\begin{align}
\label{eq:disp}
\omega(k) = \sqrt{\left(\omega_H + l_\text{ex}^2\omega_M k^2 \right) \left(\omega_H + l_\text{ex}^2\omega_M k^2 +\omega_M f(k) \right)},
\end{align}
where $\omega_H= \mu_0 \gamma (H - M)$, $\omega_M = \mu_0\gamma M$, and $f(k) = 1- \frac{1-\exp(-k s)}{k s} $. Here, $\mu_0$ is the permeability of free space, $\gamma$ is the gyromagnetic ratio, $H$ is the applied external magnetic field, $M$ is the saturation magnetization, $s$ is the film thickness and $l_\text{ex}=\sqrt{\frac{2A_\text{ex}}{\mu_0 M_0^2}}$ is the exchange length, where $A_\text{ex} = 0.4\times10^{-11}$ J/m is the exchange constant. In this paper, we will use the following values outside of the lens for the magnetization, magnetic field and film thickness, respectively: $M_0=140$kA/m, $\mu_0 H_0=200$mT, and $s_0=10\mu$m. The resulting value of exchange length is $l_\text{ex} \approx 18$ nm. These values determine $k_\text{ref}$, and thus the index will be 1 when $M=M_0$, $H=H_0$ and $s=s_0$.
\begin{figure}
\centering
\includegraphics[width=\linewidth]{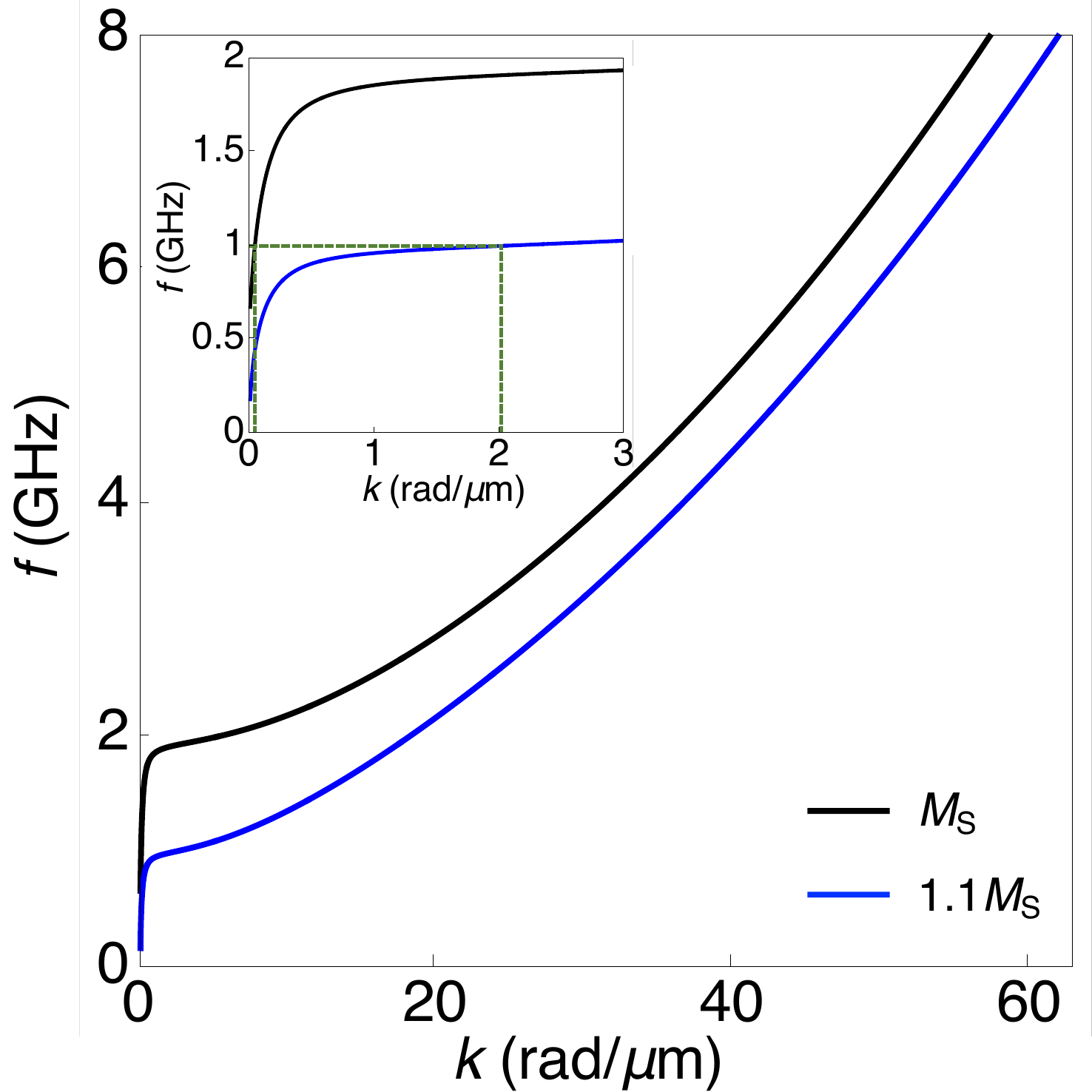}
\caption{\label{fig:disp} Dispersion relation $f(k)$ for dipole-exchange spin waves, with a zoom of the dipolar-dominated region shown in the inset. The curves use $M=M_0=140$ kA/m (black) or $M=1.1 M_0=154$ kA/m (blue). The green dashed line in the inset indicates how much the wave number changes for a fixed frequency of 1 GHz.}
\end{figure}

Using the material parameters listed above, the dipole-exchange dispersion relation is plotted in Fig. \ref{fig:disp}, where we also show the effect of increasing $M$ by 10\%. This small change in $M$ leads to a large change in $k$, and thus $n$, due to the shallow gradient in the crossover region between the dipolar-dominated and exchange-dominated regimes. The corresponding change in the index is from 1 to 54 for a fixed frequency of 1 GHz. The use of a thick film of 10$\mu$m enables a particularly large index to be achieved, because the shallow gradient extends to larger $k$ values. In comparison, a thinner film of 2 $\mu$m leads to an index change from 1 to 28 for the same 10\% increase in $M$. Note that the value of $f$ at $k=0$ marks the lower threshold of the spin wave `manifold', which corresponds to the ferromagnetic resonance frequency. In Eq. \eqref{eq:disp}, this occurs when $H=M$ and thus $\omega_H=0$.
\begin{figure*}
\centering
\includegraphics[width=\linewidth]{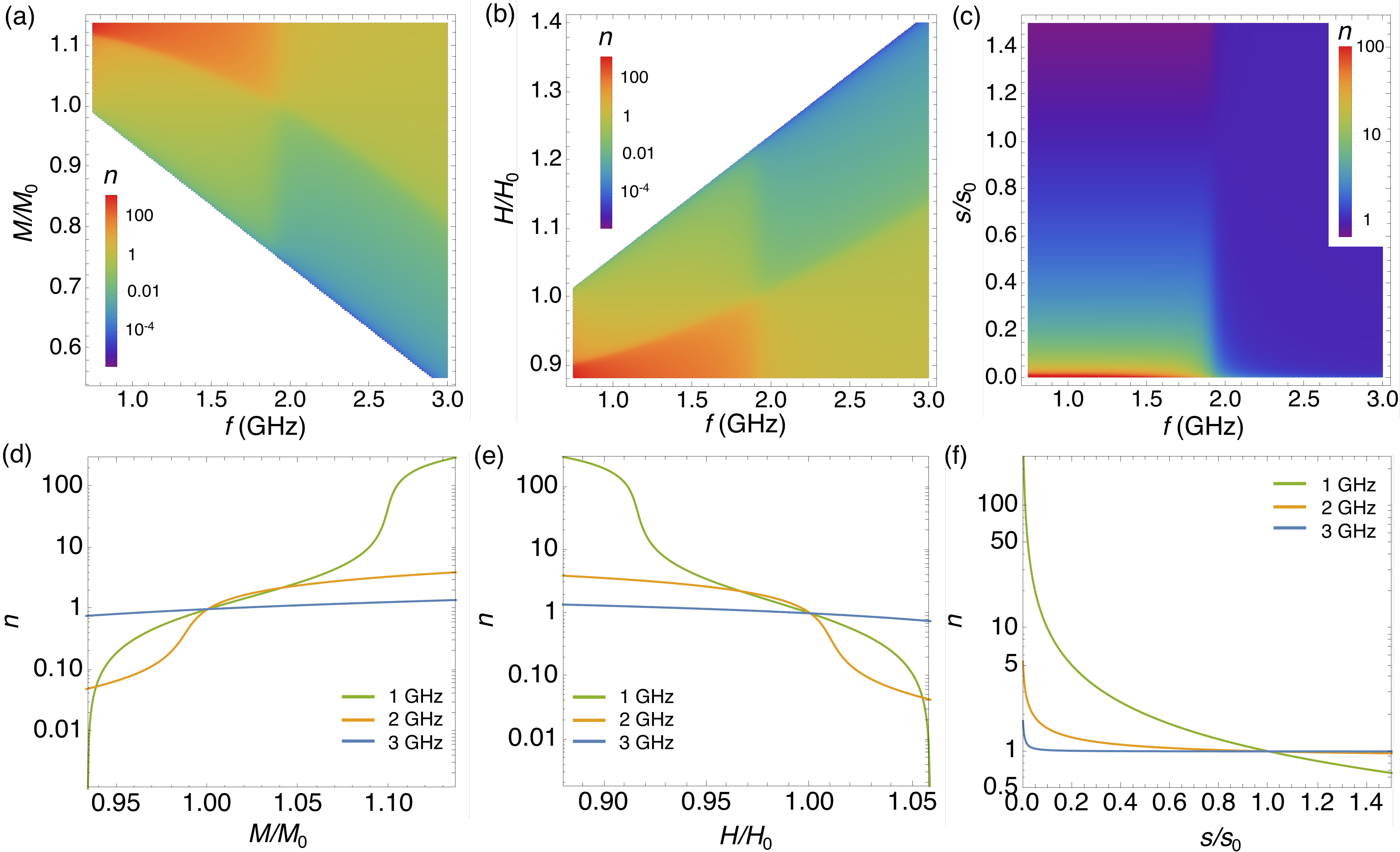}
\caption{\label{fig:indexpara} The dependence of the magnonic refractive index on (a, d) magnetization, (b, e) magnetic field and (c, f) film thickness. In (d)-(f), this dependence is shown for waves with frequencies of 1 GHz (green), 2 GHz (orange) and 3 GHz (blue). In (a)-(c), the color scale is logarithmic, along with the $n$ axis in (d)-(f) for clarity.}
\end{figure*}

In Fig. \ref{fig:indexpara}, we show how the index $n$ depends on the three parameters that can be varied in Eq. \eqref{eq:disp} ($M$, $H$ and $s$), for different incident wave frequencies $f$. In Fig. \ref{fig:indexpara} (a)-(c), we can see the distinct dipole-dominated and exchange-dominated regimes, for $f\lesssim 1.9$ GHz and $f\gtrsim 1.9$ GHz, respectively. The transition region between these two regimes is where the dispersion curve flattens at around 1.9 GHz, as we saw in Fig. \ref{fig:disp}. The white regions in panels (a) and (b) correspond to values of $M$ or $H$ for which there are no spin wave solutions for the given value of frequency, i.e. when the bottom of the spin wave manifold is above the chosen value of frequency. We have limited $M/M_0$ in (a) and (d) and $H/H_0$ in (b) and (e) to ensure that $M\leq H$, i.e. to keep the internal magnetic field positive and thus avoid any instability. In addition, we have chosen the smallest value of $f$ to correspond to a maximum wavelength of 1mm when the index is equal to 1.

Notice from Fig. \ref{fig:indexpara} that an increase in the magnetization or a decrease in the magnetic field / thickness is required to increase the index. For the former case, this may be achieved via cooling (as heating naturally reduces the magnetization \cite{vogel_optically_2015}) or doping \cite{fassbender_control_2006}. Although we show the variation of the index with thickness according to \eqref{eq:disp}, a graded index profile created in this way may induce complicated static or dynamic demagnetizing fields \cite{dove_demagnetizing_1967, schlomann_surfaceroughnessinduced_1970, langer_role_2017}, not accounted for here. However, these effects may be reduced by slowly changing the thickness over a large distance.

Strikingly, Fig. \ref{fig:indexpara} shows that just a relatively small change in $M$ or $H$ is required to produce a dramatic change in the index in the dipolar regime. This regime is therefore ideal to create the extreme refractive index profiles required for the steering lenses. In addition, refractive index profiles that require only a small change in the index, such as the Luneburg lens, may be created in this geometry by a tiny change in the same parameters \cite{whitehead_luneburg_2018}. Exchange-dominated spin waves require a large change in one of the parameters for a comparatively modest change in $n$, as we show for the 3 GHz wave frequencies in Fig. \ref{fig:indexpara}. 

\begin{figure}
\centering
\includegraphics[width=\linewidth]{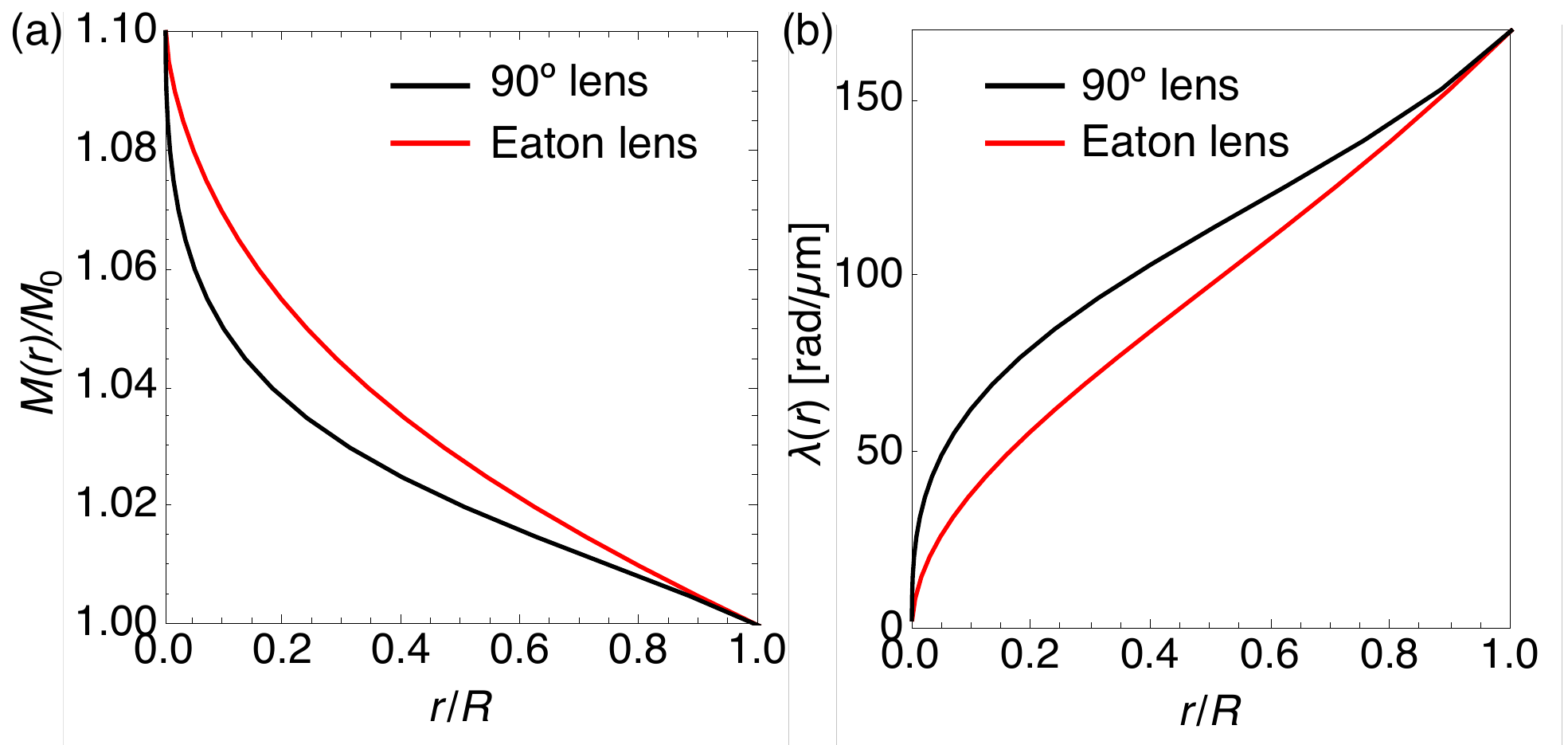}
\caption{\label{fig:profiles} (a) Magnetization profiles and (b) value of the wavelength along the radius of each lens, for the 90$^\circ$ (black) and Eaton (red) lenses. This is valid for an incident wave frequency of 1 GHz, and other parameters listed in the text.}
\end{figure}

Changing the saturation magnetization is more straightforward in micromagnetic modelling, so we will vary $M$ in this work to vary the index. Using the steering lens profiles \eqref{eq:90}-\eqref{eq:Eaton}, along with the dipole-exchange dispersion relation \eqref{eq:disp}, we can establish numerically the magnetization profile to create each lens. For the choice of material / incident wave parameters listed below, we show the required magnetization profiles in Fig. \ref{fig:profiles} (a), and the corresponding wavelength profiles in (b). For clarity, we show the profiles up to $M=1.1M_0$, which corresponds to a value of $r/R$ of $1\times10^{-5}$ and $7\times10^{-4}$ for the 90$^\circ$ and Eaton lenses, respectively. So, the majority of the profile is shown except for the singular index region in the very center.

\section{Micromagnetic Modelling}
\begin{figure}
\centering
\includegraphics[width=\linewidth]{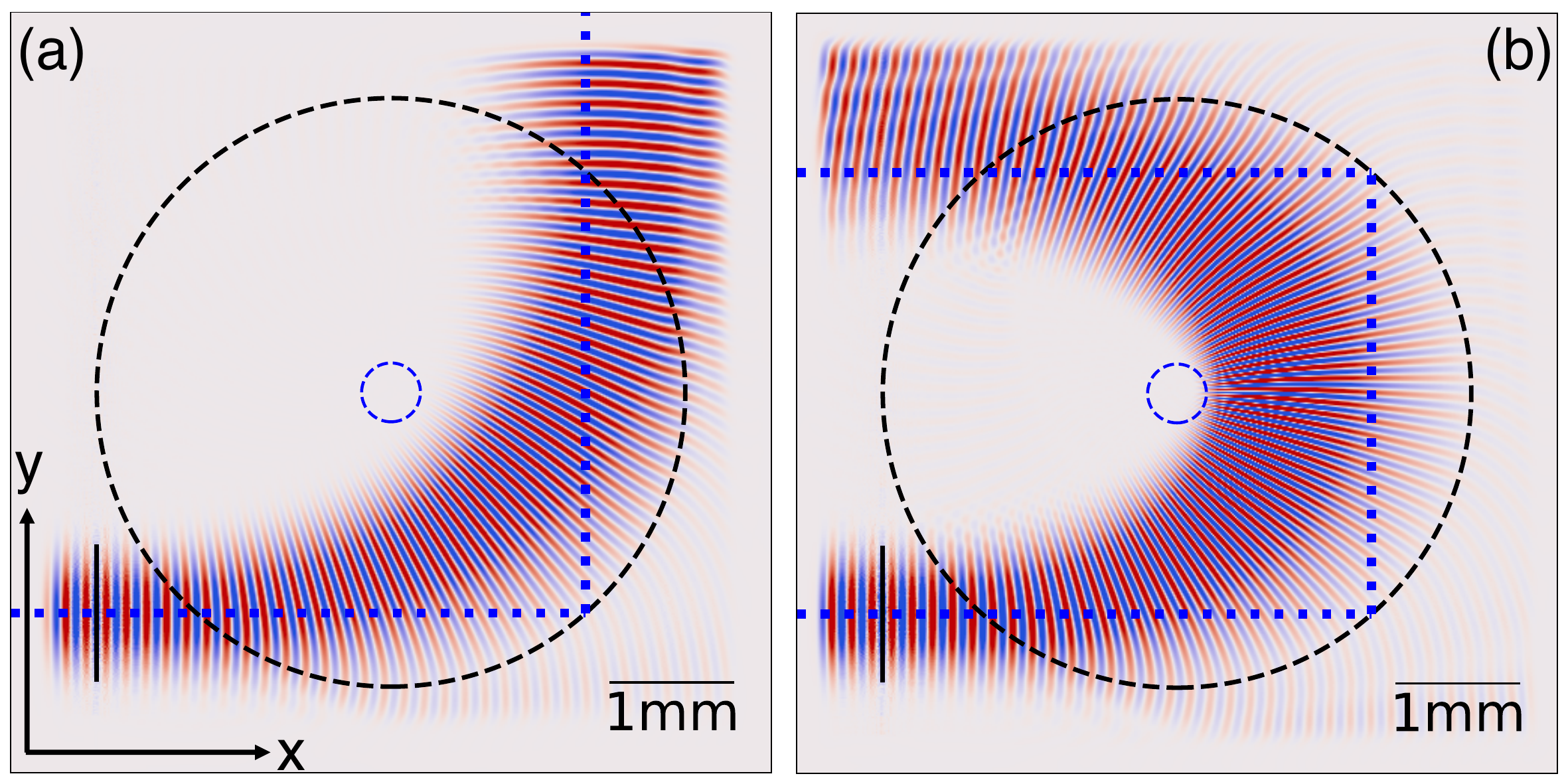}
\caption{\label{fig:beams} Steady-state snapshot of beams travelling through the (a) 90$^\circ$ and (b) Eaton lens, sized at $r=14\lambda$. The inner dashed circle indicates $r/R=0.1$, and blue dotted guide lines are shown to indicate the $90^\circ / 180^\circ$ angles. The black line on the lower left in each image indicates the source region for the beam. }
\end{figure}

In order to verify the above analysis, we have performed micromagnetic simulations using MuMax3 software \cite{vansteenkiste_design_2014-1}. We model an yittrium iron garnet (YIG) film with thickness $s_0=10\mu$m in the $z$ direction, and extent of around 6mm$\times$6mm in the $x-y$ plane. The $x-y$ axes are defined in Fig. \ref{fig:beams}. As before, the saturation magnetization outside each lens is set to $M_0=140$kA/m and the bias magnetic field is $\mu_0 H_0=200$mT in the $z$ direction. We set the Gilbert damping parameter to $\alpha = 1\times10^{-4}$.

We use a cell size of $1.5\times1.5\times10\mu$m with $4096\times4096\times1$ cells in the $(x,y,z)$ directions. This choice of cell size is a compromise between resolving the smallest possible wavelength, and being able to represent a large enough lens. From Fig. \ref{fig:profiles} (b), we can see that if we direct the incident waves to avoid the region $r/R<0.1$, then the smallest wavelength should easily be greater than 15$\mu$m, which is 10 times larger than the cell size in the film plane. This approach is a necessity for the modelling, but should not be a limitation for any future experiments. If the profile in Fig. \ref{fig:profiles} can be created, this should represent the refractive index profile almost exactly. 

We now describe the form of the incident wave. The lenses are primarily designed to steer a collimated beam, and we create this with a magnetic field of the form $[1 - \exp(-0.1\omega_0t)]\sin(\omega_0 t)$ in time, where $\omega_0 = 2\pi f_0$ and $f_0$ is the excitation frequency. Spatially, this magnetic field is Gaussian in $y$ and has a step profile in $x$, 8 cells wide, similar to the approach in Ref. \onlinecite{gruszecki_goos-hanchen_2014}. The magnetic field is directed along $x$, with an amplitude of 0.2 mT, and a frequency of $f_0= 1$ GHz. We also find that the lenses work well with a wave packet, which we position to be partly steered by the lens and partly unaffected by it, similar to the approach in Ref. \onlinecite{ma_omnidirectional_2009}. We create the wave packet by amending the beam's magnetic field profile to be of the form $G[x]G[y]G[t]\sin(\omega_0 t - k_0 x)$, where $G[x,y,t]$ is a Gaussian in $x$, $y$ or $t$, $k_0=2\pi/\lambda_0$, and $\lambda_0\approx170\mu$m is the spin wave wavelength outside of the lens for excitation frequency $f_0$. We also employ absorbing boundary layers along the edges in the $x$ and $y$ directions \cite{venkat_absorbing_2018}.

A perfectly graded index is not possible in finite difference simulations, but a stepped profile can work effectively if the steps are much smaller than the wavelength. This also holds true in experiments, as per the metamaterial approach \cite{vasic_controlling_2010}. In the model, we allocate 233 concentric circular regions to the lens, where the radius of each region is sized to ensure that $M$ steps up by equal amounts each time, until reaching $1.1M_0$ as per Fig. \ref{fig:profiles}. However, this profile will not be matched exactly due to the cell size, especially where $M$ is required to change substantially on a length scale which is smaller than the cell size. This is only an issue towards the center of the lens and may lead to some scattering, which should be mitigated somewhat by avoiding the central region of $r/R<0.1$.

In Fig. \ref{fig:beams}, we show the beam's trajectory through each lens, after a long enough time has elapsed. Both lenses are sized at $R=14\lambda_0$, to ensure the beam is mostly contained within the lens. We can see that the 90$^\circ$ lens works particularly well to bend the beam by the required angle, although there is some expected spreading of the beam within and on exiting the lens, making it difficult to see if the trajectory follows the required angle exactly. The Eaton lens is quite sensitive to the placement of the beam, as the beam tends to spread into the central region, where the cell size limits how well we can represent the refractive index profile. We have found that positioning the beam towards the edge of the lens means the central region is mostly avoided, and the trajectory is around 175$^\circ$. Note for both images that the absorbing boundaries have absorbed the edge of the outgoing beam, so the actual outgoing beam is a little wider than shown.
\begin{figure}
\centering
\includegraphics[width=\linewidth]{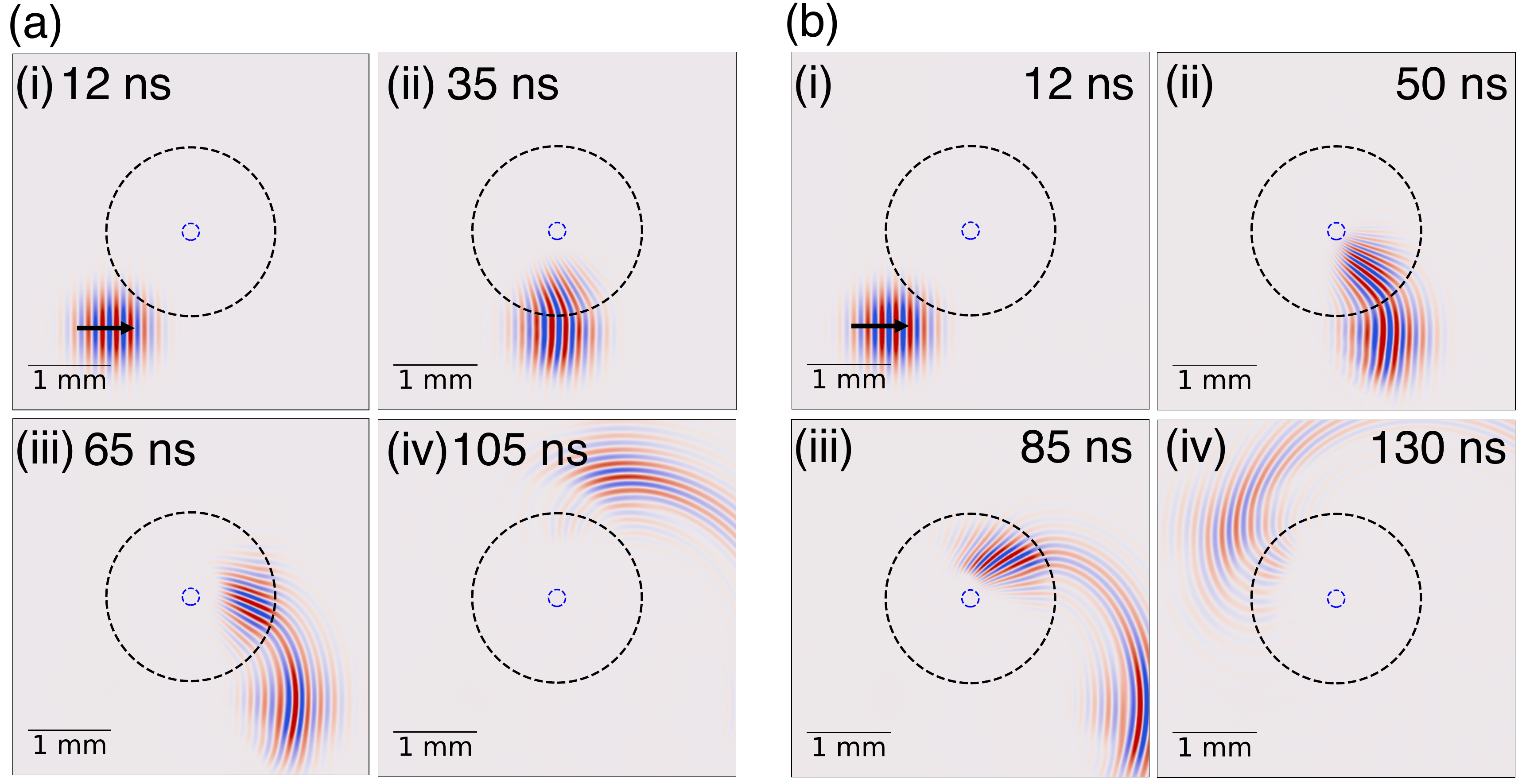}
\caption{\label{fig:90_180} Consecutive snapshots of the wave packet moving through the (a) 90$^\circ$ and (b) Eaton lenses with $R=6\lambda$, shown from (i)-(iv). The $m_x$ component is shown, saturated for clarity. The inner circle indicates $r/R=0.1$. The black arrow in panels (i) indicates the initial propagation direction of the wave packet.}
\end{figure}

We show the results for the wave packet incident on the 90$^\circ$ and Eaton lenses in Fig. \ref{fig:90_180}, and the corresponding videos are provided in the Supplemental Material \footnote{See Supplemental Material for videos of the results shown in Fig. \ref{fig:90_180}}. We can see that in each case, the portion of the wave packet that enters the lens is steered approximately by the required angle, and remains `connected' to the other portion of the packet that does not enter the lens and hence continues on the original trajectory. Interestingly, this implies that the part of the wave joining these two portions of the wave packet experiences an effective graded index, despite being in a homogeneous medium; its wavefronts must be curved, to bridge the two diverging parts of the wave packet \cite{philbin_making_2014}. The use of the lenses in this way is similar to a beam divider, and may be a way to send different portions of the same wave (beam or packet) to more than one output, albeit with some loss en route.
 \begin{figure}
 \centering
 \includegraphics[width=\linewidth]{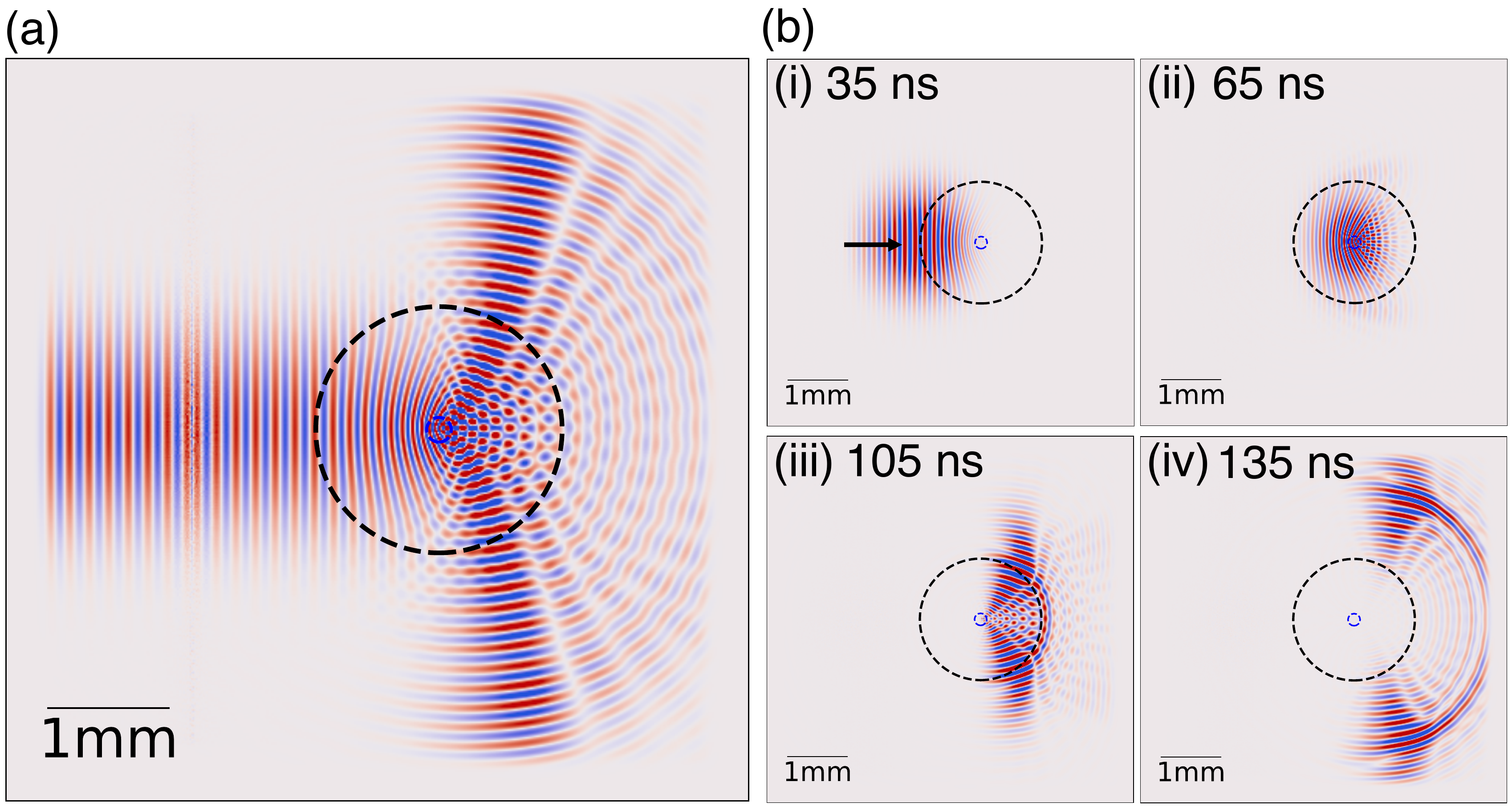}
\caption{\label{fig:beamdivider} (a) Steady state snapshot of the beam and (b) snapshots in time of the wave packet moving through the 90$^\circ$ lens ($R=6\lambda$) to demonstrate its use as a beam divider. The $m_x$ component is shown, saturated for clarity. The inner circle indicates $r/R=0.1$.}
\end{figure}

In Fig. \ref{fig:beamdivider}, we show another use for the 90$^\circ$ lens when the beam is instead positioned to enter the lens symmetrically about the center. In this case, the lens acts as a $\pm90^\circ$ half-power beam divider, proposed by Ref. \onlinecite{cornbleet_generalised_1981}. This works well for both a beam (Fig. \ref{fig:beamdivider} (a)) and wave packet (Fig. \ref{fig:beamdivider} (b)), albeit with some scattering from the central region. Note that we have broadened the excitation width across the $y$ direction, to ensure that the beam is exposed to as much of the lens as possible, without having to reduce the lens size. In addition, we have reduced the excitation amplitude to 1 mT in both cases, to avoid a nonlinear response when the wave encounters the high-index central region. The corresponding videos are provided in the Supplemental Material \footnote{See Supplemental Material for videos of the results shown in Fig. \ref{fig:beamdivider}}.
 \begin{figure}
 \includegraphics[width=\linewidth]{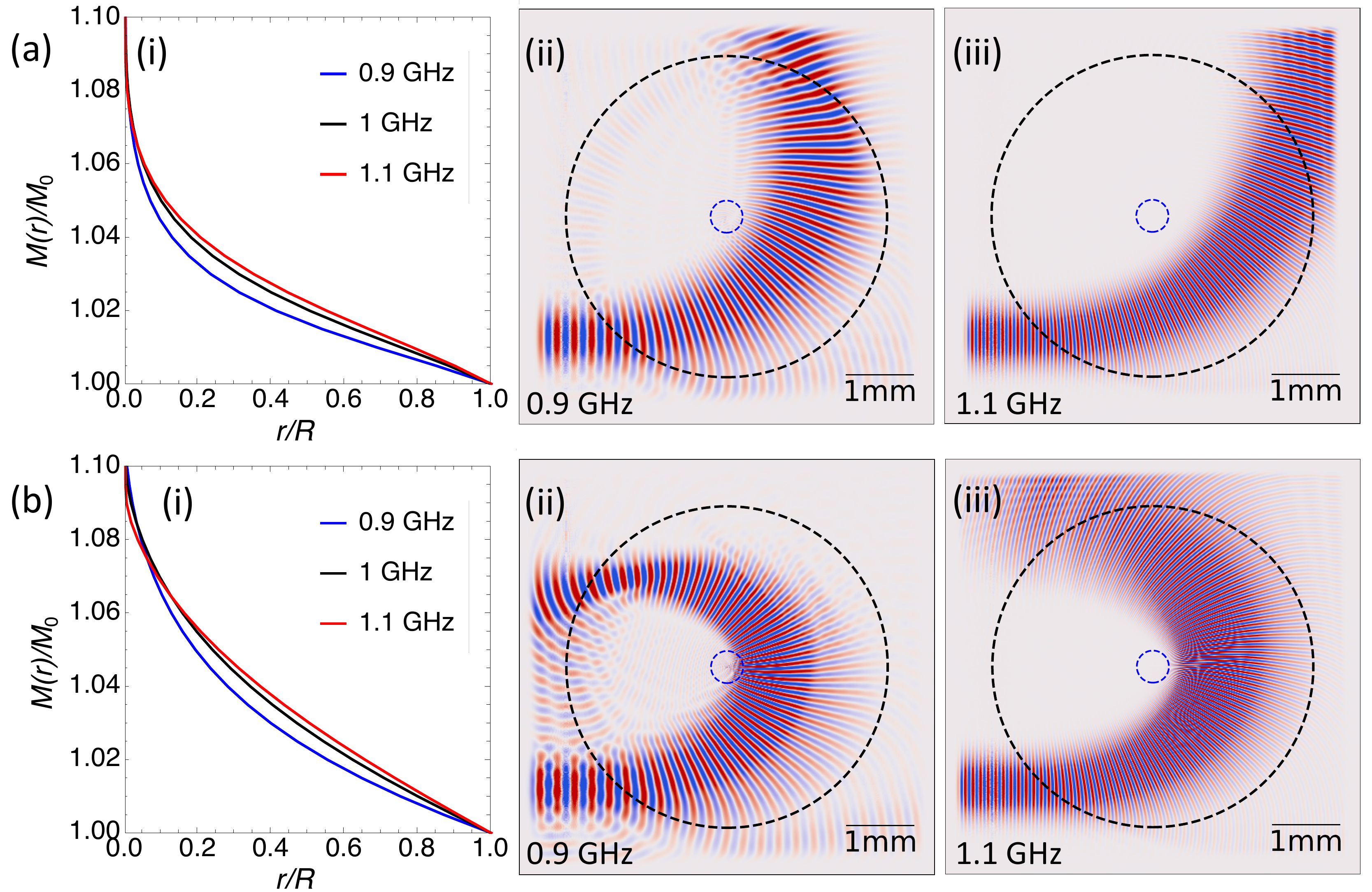}
\caption{\label{fig:diff_freq} Demonstration of the effectiveness of the (a) 90$^\circ$ and (b) Eaton lenses from Fig. \ref{fig:beams}, for different incident wave frequencies (ii) $f=0.9f_0$ and (iii) $f=1.1f_0$, with $f_0=1$ GHz. For comparison, the magnetization profiles which would be required to make the lenses for each frequency is shown in (i). The inner circle indicates $r/R=0.1$.}
\end{figure}

So far, we have seen the results for each lens when the refractive index profile is designed correctly for the incident wave. However, we would now like to demonstrate that the lenses still work reasonably well when the incident wave frequency is slightly different from the optimal value. As we will see, this is equivalent to designing a slightly incorrect magnetization profile for a certain choice of frequency. In Fig. \ref{fig:diff_freq}, we change the frequency of the incident wave by $\pm 10\%$ from $f_0 = $1 GHz, and these waves travel along the profile designed to work for an incident wave frequency of $f_0$, for the 90$^\circ$ lens in Fig. \ref{fig:diff_freq} (a) and Eaton lens in Fig. \ref{fig:diff_freq} (b). In panel (i), we show the magnetization profiles that would be required to make the lenses for each frequency. We then show the results for the 0.9 GHz and the 1.1 GHz beams in panels (ii) and (iii), respectively. Recall that these 0.9 GHz and 1.1 GHz beams should rotate by 90$^\circ$/180$^\circ$ only when they encounter their respective magnetization profiles in panel (i), but they are instead traversing the profile designed for the 1 GHz wave. As a result, we see that the 0.9 GHz beam rotates too much, and the 1.1 GHz beam does not rotate enough in each case. The angles are again difficult to quantify exactly due to the beam spreading, but are around 10$^\circ$-20$^\circ$ away from the target angle in each case. This suggests that if the wave trajectory is not quite right, then the correct trajectory may be recovered by adjusting the wave frequency accordingly.

\section{Conclusions}

In summary, we have demonstrated how steering lenses with singular graded index profiles can be almost exactly realized for spin waves with a 10\% change in either the external magnetic field or magnetization, in the dipole-dominated regime. We have shown the operation of two such lenses in micromagnetic modelling by changing the magnetization, but the theory is applicable for rotation by any angle, from any angle of incidence. As long as the index is smoothly graded, the lenses should be robust to small deviations in the profile, and small deviations in rotation angle may be corrected by changing the incident wave frequency. Our results demonstrate the potential of magnonics for realising extreme ranges of the refractive index, something that is far more difficult to achieve in other areas of wave physics.

\begin{acknowledgments}
This research has received funding from the Engineering and Physical Sciences Research Council (EPSRC) of the United Kingdom, via the EPSRC Centre for Doctoral Training in Metamaterials (Grant No. EP/L015331/1). SARH would like to thank the Royal Society and TATA for financial support (Grant No. RPG-2016-186).
\end{acknowledgments}



%



\pagebreak

\clearpage

\pagebreak

\setcounter{equation}{0}
\setcounter{figure}{0}
\setcounter{page}{1} 
\renewcommand{\thefigure}{S\arabic{figure}} 
\renewcommand{\theHfigure}{Supplement.\thefigure} 

\onecolumngrid 

\section*{Supplemental Material}
\subsection*{List of Supplementary Animations and their Captions}

Animation 1 (a) - Wave packet incident on 90 degree lens: Video corresponding to Fig. 6 (a) in the main text, showing the $m_x$ component of the wave packet moving through the 90$^\circ$ lens.  \\

Animation 1 (b) - Wave packet incident on 180 degree lens: Video corresponding to Fig. 6 (b) in the main text, showing the $m_x$ component of the wave packet moving through the Eaton (180$^\circ$) lens. \\

Animation 2 (a) -  90 degree lens as a beam divider: Video corresponding to Fig. 7 (a) in the main text, showing the 90$^\circ$ lens acting as a $\pm90^\circ$ half-power beam divider. The $m_x$ component of the beam is shown. \\

Animation 2 (b) - 90 degree lens as a wave packet divider: Video corresponding to Fig. 7 (b) in the main text, showing the 90$^\circ$ lens acting as a $\pm90^\circ$ half-power beam divider for an incoming wave packet. The $m_x$ component of the wave packet is shown.  \\

Each animation uses the parameters stated in the main text.

\clearpage


\end{document}